\documentstyle[12pt,a4,epsfig]{article}

\begin{document}
\sloppy
\thispagestyle{empty}

\mbox{}

\hskip 10.60truecm FUB-HEP/95-15
\vspace*{\fill}
\begin{center}
{\LARGE\bf Theoretical aspects of singly polarized
hadron-hadron collisions$^{\S ~*}$}\\

\vspace{2em}
\large
Meng Ta-chung
\\
\vspace{2em}
{\it  Institut f\"ur Theoretische Physik, Freie Universit\"at Berlin}

{\it Arnimallee 14, 14195 Berlin, Germany}

\end{center}
\vspace*{\fill}
\begin{abstract}
\noindent
The special role played by singly polarized
high-energy hadron-hadron collisions
in Spin Physics is discussed:
In such processes, the measured and the
calculated quantities can be and have been directly
compared with each other ---
without data-extrapolation and without sum rules.
It is in this kind of processes,
where significant asymmetries (up to
30-40\%) have been  observed.
It is also in this kind of processes,
where the obtained data and the predictions
of the conventional theories dramatically
disagree with each other. Attempts to understand
the existing data are briefly
summarized. Predictions for further experiments are
presented.

\end{abstract}
\vspace*{\fill}

\vskip 2.0truecm

\noindent
------------------------------------------------

$\S$Invited talk given at the Workshop on the Prospects of Spin Physics at

\hskip 0.5truecm HERA, DESY Zeuthen, August 28-31, 1995.

$^*$Supported in part by Deutsche Forschungsgemeinschaft
DFG: Me 470/7-1.
\newpage
%
\section{Introduction}
\label{sect1}

In this talk I discuss high-energy hadron-hadron collisions in which either
the projectile or the target is polarized transversely with respect to the
scattering plane. A number of experiments[1-10] of this kind have been
performed in the past, among which the
elastic proton-proton scattering[1]
using polarized target and the inclusive pion-production[2-9]
using polarized proton and antiproton beams are probably
the most well-known
ones --- in and outside the Spin Physics Community.

This kind of spin-dependent collision processes is of particular interest
for the following
reasons:

(a) It is conceptionally simple !

(b) Here, the measured and the calculated quantities,
for example the left-right-asymmetries in inclusive pion production,
can be and have been directly compared with one another
--- without data-extrapolation and without  sum rules.

(c) A considerable amount of high-energy
single-spin-asymmetry data
(up to 200 GeV/c incident momentum in laboratory
for inclusive meson- and direct-photon production)
are now available, and extremely striking features
have been observed. Further experiments
of this kind will be performed at
higher energies --- at RHIC, UNK and
perhaps also at HERA.

(d) The existing data[1-10] drastically
disagree with the theoretical expectations[11]
made on the basis of usual (leading twist)
perturbative QCD and/or on conventional
pQCD-based hard-scattering  models.
This seems to suggest that mechanisms beyond the
usual pQCD may play a significant role in such processes.

This talk will be devided into the following parts.
After this introduction,
I shall first briefly remind you of the characteristic
features of the existing data which are so striking!
Then I shall compare these features with
the expectations of pQCD and the usual
pQCD-based hard-scattering models.
After this,
I shall discuss some non-perturbative aspects
--- in particular  a simple relativistic quark model.

\section {Characteristic features of the existing single-spin
asymmetry data}

In $p+p(\uparrow)\to p+p$[1],
it is observed that
the analyzing power $A$ is significantly different from zero when the
transverse momentum $(p_\bot)$ of the scattered  protons is
large [$p_{\perp}^2 > 5$ (GeV/c)$^2$ say], and $A$ increases
with increasing $p_\bot^2$.

In $p(\uparrow)+ p\to (\pi^0,\pi^+,\pi^-,\eta,\gamma_{dir})+ X$ and
$\bar p(\uparrow) + p\to (\pi^0,\pi^+,\pi^-)+ X$ [4-10],
it is seen that the left-right asymmetry
$A_N$ has the following properties:

First, $A_N$ depends strongly on $x_F$ (the Feynman $x$-variable),
but only very weakly (if at all) on $p_\perp $
(the transverse momentum of the observed particle).
To be more precise: $A_N$ is consistent with zero near $x_F=0$
(the central rapidity region), independent of $p_\perp $[9,10].
But, it becomes nonzero at  about $x_F=0.4$, increases
monotonically and reaches up to
$40\%$ near $x_F=0.8$.
In other words, the observed[4-8]
asymmetry is significantly non-zero in the
projectile-fragmentation region.
In this kinematical region
the asymmetries in the event sample with
$p_\perp >0.7$ GeV/c are somewhat larger
in magnitude than those in the  $p_\perp <0.7$GeV/c event-sample.

Second, the observed asymmetries[4-8] $A_N$ for $\pi^0,\pi^+$ and
$\pi^-$ are very much different from one another.
For example, in
$p(\uparrow) + p\to (\pi^0,\pi^+$ or $\pi^-)+X,
A_N(\pi^+)>A_N(\pi^0)> 0$  but $A_N(\pi^-)<0$.
That is, the observed left-right asymmetry
$A_N$ is flavor-dependent~!

Third, the observed asymmetries depend on the projectile.
It is seen[8] in $\bar p(\nobreak\uparrow\nobreak)
+p\to (\pi^0,\pi^+$ or $\pi^-$)
that $A_N(\pi^+)<0,A_N(\pi^0) > 0$ and $A_N(\pi^-) >0$!

Fourth, it has been reported[3]
that the left-right asymmetries
in the projectile fragmentation region in $\pi^-+
p(\uparrow) \to (\pi^0$ or $\eta)+X$ are consistent with zero.

\section {Perturbative QCD and pQCD-based hard-scattering models}

What are the expectations of pQCD and the usual
pQCD-based hard-scattering models?
The relationship between QCD and the polarization of scattered or produced
quarks has been discussed already in the late 1970's by G.L.~Kane, J.~Pumplin
and W.~Repko[11]. They pointed out that the polarization of scattered or
produced quarks in large-$p_\bot$ hadron-hadron collisions can be calculated
in pQCD, and according to the usual pQCD
calculations the predicted value should be zero
(This is because the contribution is proportional to
$m_q/\sqrt{s}$, where $m_q$ is the quark mass and
$\sqrt {s}$ is the total cms energy
which is much much larger than $m_q$. Note also that for $m_q\to 0$,
there is no helicity flip in the Born diagram or box diagram).
Since the individual quark-quark scattering
in general produce only a small left-right asymmetry,
pQCD necessarily predicts a small left-right asymmetry, independent of the
details of the wave function of the quarks in polarized nucleon.
These statements apply to exclusive as well as to inclusive processes.
Note that since the latter is much easier to describe theoretically, we
shall confine our discussions in this talk
on inclusive production processes only.

PQCD-based hard-scattering models have been
discussed by many authors[11-18].
In this kind of description, the cross section for the
inclusive production of large $p_\perp $ pions in
$p(\uparrow )+p \to \pi + X$ can be
expressed as a convolution of the
elementary cross-section
(which describes the scattering of quarks and gluons),
the number-densities of the quarks and gluons
inside the polarized and the
unpolarized protons (spin-dependent and spin-averaged structure
functions), and the number-densities of pions in quarks/gluons
(fragmentation functions).
While the elementary cross sections are calculable in pQCD
(provided that the corresponding running coupling constant
$\alpha_s$ is sufficiently small),
the structure functions and the fragmentation functions
are {\it not} !
The reason is:
 pQCD is no more valid for soft-processes where
long-distance color-interactions play the dominating role.
Since leading twist pQCD gives negligible (essentially zero)
contribution,
it is clear that in order to obtain a non-zero single-spin
asymmetry, higher twists must be included in calculating
the elementary cross sections, and
spin effects need to be introduced
into the structure
functions and/or the fragmentation functions.
Now, let me show you a few
explicit examples of such models.
In a recent paper[18], M.~Anselmino, M.~Boglione and F.~Murgia
reproduced the $p(\uparrow )+p \to (\pi^+,~\pi^0,~\pi^-)+X$ data
in a model of this type. They assume that the factorization theorem
holds in the helicity basis for higher twist contributions and they
assume that
non-perturbative and intrinsic transverse momentum
effects can be properly taken into account
in a phenomenological approach.
In their model[18], $A_N$ depends on a set of 6 free parameters
which appear in  spin-dependent structure function $I_{+-}^{a/p}$.
The flavor-dependence of $A_N$ for $\pi^+,~\pi^0$ and $\pi^-$
is reproduced by choosing different sets of
parameters for the $u$- and the $d$-quarks.
The authors stressed that they have no problems with
the time reversal invariance of QCD, because
they consider higher twists and they do not exclude
soft initial state interactions.
The problem of flavor-dependence has also been addressed by
A.V.~Efremov, V.M.~Korotkiyan and O.V.~Teryaev[17].
They calculated the
single-spin parton asymmetries for high $p_\perp $ gluon
and quark production using the sum rules for
the twist-3 quark-gluon
correlators and the twist-2 distribution function.
It is reported in their paper[17] that
the difference in sign of
asymmetries for $\pi ^+$, $\pi ^0$ and $\pi ^-$
production can be reproduced by inserting the
empirical values for
the spin-contents of the $u$- and $d$-quarks
$\Delta u=0.80\pm 0.04$ and $\Delta d=-0.46\pm 0.04$
from the polarized lepton-nucleon scattering data
are different in sign.
Furthermore, higher order elementary interactions and
higher twist distribution functions
have been used by J.~Qiu and G.~Sterman[14], and by
A.~Sch\"afer, L.~Mankiewicz, P.~Gornicki
and S.~G\"ullenstern[15], where
non-zero single-spin
asymmetries in $p(\uparrow )+p\to \gamma _{dir}+X$ at large
$p_\perp $ have  been obtained.
Since more about pQCD-based hadron-scattering models will also
be discussed by Dr. Teryaev, the next speaker, I shall
now change my subject.

\section {A non-pQCD approach}

A different approach has been
pursued by the FU-Berlin group.
Instead of performing calculations
in pQCD or in the  framework of
pQCD-based hard-scattering  models, they carried out a
systematic analysis of all available data
directly or indirectly
related to the singly
polarized hadron-hadron collision processes.
They observed that the characteristic features of the
asymmetry data have much
in common with the typical properties of soft
hadronic processes in the fragmentation regions of unpolarized
hadron-hadron collisions.
Since this similarity strongly suggests
that a considerable part of the mechanisms
which are responsible for the observed asymmetries
are non-pQCD in nature, they decided to try a non-perturbative approach.
In this connection,
it is important to note the following:

(I.) Inclusive meson-production
in hadron-hadron collision with unpolarized projectile
and target have been extensively studied
already in the 1960's and 1970's, where in
particular, leading particle effect,
limiting fragmentation of the  projectile
have been observed[19] in the kinematical region $x_F\ge 0.4$:
In term of quarks, these experimental
facts strongly suggest that
valence quarks play a dominating role in this
kinematical region.
In fact, it has been shown[20-24] that
part of the mesons observed
in this region are due to direct
formation (fusion) of the valence quarks
of the projectile P and antiquark from
the sea of the target T.
A natural question that can and should be asked is:
Should this be completely different
when the {\it projectile is polarized} ?"

(II.) $A_N\not= 0$ for $x_F\ge 0.4$ means:
Transverse motion of the produced meson due to
the transversely polarized projectile
$P(\uparrow)$ and hence that of
of the valence quark in $P(\uparrow)$ is asymmetric.
Since this occurs in, and only in, the fragmentation region of the
projectile hadron, it  is natural to ask:
Can this be due to the valence quarks in
the transversely polarized projectile ?
The answer is ``Yes !"; and the reason is :
A valence quark can be considered as
a Dirac-particle in an effective confining
potential (due to the existence of other constituents).
Hence, the quantum numbers which characterizes
the eigenstates of such a valence quark with given color and flavor are:
$(\epsilon~,~j~,~j_z,~P)$ and in particular
$(\epsilon_0~, 1/2~,~ \pm 1/2~,~ +1)$ for the ground state.
Note that $j$ is the total angular momentum
(not the orbital angular momentum because the latter is
{\it not} a good quantum number) !

(III.) It is known that baryon's magnetic moments can be well-described
in terms of those of the quarks. In this connection
baryons's wave functions can be readily constructed
not only in the static quark-model[25], but also in a relativistic
quark model. This can be done simply by replacing
in the static quark model[25], the Pauli 2-spinors
by the corresponding Dirac 4-spinors.
In terms of the quark-magnetic-moments,
the resulting formulae for the magnetic
moments of the baryons in the relativistic quark-model
have exactly the same form -- independent of the confining potentials.
Hence, this baryon-wavefunction describes the baryon-magnetic moments as
good (or as bad) as the static quark model[25].
The same wave functions can be used to determine the polarization of the
valence quarks in the polarized baryons.
In particular, the proton, on the average,[26,27]

$5\over 3$ of the 2~~$u$-valence
quarks are in the same direction

$1\over 3$ of the 2~$u$-valence quarks are in the opposite direction

$1\over 3$ of the 1~$d$-valence quark is in the same direction

$2\over 3$ of the 1~$d$-valence quark is in the opposite direction\\
{}~\\
This means: There is asymmetry in valence quark polarization;
and this asymmetry
is flavor-dependent!!
Furthermore, since the wave function of any antibaryon
can be readily obtained from the corresponding
baryon wave function, the relationship between the results in
$p(\uparrow ) +p$ and $\bar p(\uparrow ) +p$ should be predictable.

(IV.) Hadrons are spatially extended objects,
and color-forces exist only inside the hadrons.
Hence, due to causality, significant surface effects
are expected in hadron-hadron collisions in general,
and hadronic inclusive
production processes in particular.
One of the immediate consequences
in such production process, in which a valence quark
of the projectile and an antiseaquark of the
target directly form a system, is the following:
Only color-singlet $q\bar q$
systems directly formed near the
front-surface can acquire extra transverse-momenta
due to the orbital motion of the valence quarks (Cf. Fig.1).

Based on these experimental facts and theoretical arguments, the
Berliners proposed a relativistic quark-model (BRQM)[26-32],
the basis of which are the following:

(I.) Part of the mesons observed in the projectile fragmentations region
$(x_F\ge 0.4)$ are directly formed by the valence quarks of the projectile
-- also when the projectile hadron is polarized.

(II.) A valence quark of a hadron can be considered as a Dirac particle
in an effective confining potential (due to the other constituents of the
hadron). Hence, the ground state of a given quark with a given color and a
given flavor can be characterized by its energy $\epsilon$,
its total angular momentum
$(j=1/2, j_z= -1/2$ or $-1/2)$) and its
parity $(P=1)$.

(III.) In a relativistic quark model,  the  wave functions
for the baryons can be obtained simply
by replacing the Pauli 2-spinors in the
static quark model by the corresponding Dirac 4-spinors which describe the
ground states of the valence quarks.

(IV) Like all hadron-hadron collisions,
 ``surface effect"  should
play  an important role also in inclusive
production processes.

These four points (I to IV) are the cornerstones of the proposed
Berliner relativistic quark model BRQM[26-32].
They agree with the existing unpolarized
hadron-hadron collision data,
because they have in fact been extracted from
experimental facts. They agree with the basic properties of QCD ---
the only candidate for hadronic interactions;
yet, they are definitely {\it beyond} the
perturbative QCD regime.
The reason is:
None of the key concepts which have been used to
describe the above-mentioned characteristic properties
--- in particular neither
leading particle effect, nor confining
potentials, nor baryon wavefunctions, nor
surface effects in hadron-hadron collisions can be
described by pQCD.

This model has  already been worked out. The calculations and the
results can be found in Refs. [26-32].
In this talk, I shall first show you
some of the results which can be
readily seen without knowing the details.

In order to compare with the
$p(\uparrow) + p \to$ meson $+ X$, and
$ \bar p(\nobreak\uparrow\nobreak) + p \to $meson $+ X$
experiments where the mesons are:
$\pi^+,\pi^-,\pi^0$, or $\eta$ the authors use
a right-handed Cartiesian coordinate system in which
the projectile is moving in the
positive $z$-direction, the polarization
``up" is in the positive $x$-direction
while the origin is fixed at the c.m.s.
of the colliding hadron-hadron system.
According to BRQM, the following are expected :

(A) In the projectile-fragmentation region of inclusive meson production
processes $p(\uparrow) +p\to (\pi^+\pi^-,\pi^0$ or $\eta)+X$ in which
the valence quarks of the upward polarized projectile proton contribute, the
produced $\pi^+,\pi^0$ and $\eta$ go left, while $\pi^-$ go right.

(B) By using transversely polarized {\it antiproton} instead of proton-beam,
$\pi^0$ and $\eta$ behave in the {\it same} way as that in the proton-beam
case, but $\pi^+$ and $\pi^-$ behave {\it differently}.
They change their roles ! (Cf. Tables 1 and 2).

(C) In the corresponding production processes using pseudoscalar meson beams
---
irrespective of what kind of target is used and whether the target is polarized
---
there should be no left-right asymmetry in the projectile fragmentation region.

(D) The asymmetry of the produced mesons is expected to be more significant for
large $x_F$ in the fragmentation region of the transversely polarized
projectile.

(E) Not only mesons but also lepton-pairs in such experiments are expected to
exhibit left-right asymmetry.

The qualitative features mentioned in
(A), (B), (C) and (D) agree well with
experiments [3-9].

The associations mentioned in (B) and (C)
have been predicted [26] before the
corresponding data[3,8] were available.
The existence of left-right asymmetry
for lepton-pairs is a further prediction,
which still need to be verified experimentally.
Qualitative predictions for other processes such as
$p(\uparrow)+p\to K+X$ and
$\bar p(\nobreak\uparrow\nobreak)+p\to K+X$
can be and have already been made[30].
It is expected in particular that,
\vspace{0.4truecm}

$A_N^{p(\uparrow)+p\to K^++X} (x_F)$ should be similar to
$A_N^{p(\uparrow)+p\to\pi^++X}(x_F)$,

$A_N^{p(\uparrow)+p\to K^0+X} (x_F)$ should be similar to
$A_N^{p(\uparrow)+p\to\pi^-+X}(x_F)$,

$A_N^{p(\uparrow)+p\to K^-+X} (x_F)=0$ (because $K^- = \bar u s)$

$A_N^{p(\uparrow)+p\to \bar K^0+X} (x_F)=0$
(because $\overline{K^0} = \bar d s)$

\vskip 0.2truecm

$A_N^{\bar p(\uparrow)+p\to K^-+X} (x_F) \approx
A_N^{p(\uparrow)+p\to K^++X}(x_F)$

$A_N^{\bar p(\uparrow)+p\to \bar K^0+X} (x_F) \approx
A_N^{p(\uparrow)+p\to K^0+X}(x_F)$

$A_N^{\bar p(\uparrow)+p\to K^+ +X} (x_F)=0$ (because $K^+ = u\bar s)$

$A_N^{\bar p(\uparrow)+p\to K^0+X} (x_F)=0$ (because $K^0 = d \bar s)$

\vskip 0.2truecm

Let me now show you some of the quantitative results.
Because of the limited time,
I shall not discuss the details,
but only show you the following:
($\alpha $) in Fig.2: comparison between data[4-7]
and the calculated result for
$p(\uparrow )+p\to (\pi^+,~\pi ^-,~\pi^0)+X$,
together with the predictions for
$p(\uparrow )+p\to l\bar l+X$, and
$\bar p(\uparrow )+p\to l\bar l+X$.
($\beta $) in Fig.3: the calculated left-right asymmetry $A_N$
as function of $x_F$ for inclusive lepton-pair production
using $\pi ^-$-beam and transversely polarized nucleon
and nuclear targets:
$\pi^-+p(\uparrow )\to l\bar l+X$,
$\pi^-+n(\uparrow )\to l\bar l+X$,  and
$\pi^-+D(\uparrow )\to l\bar l+X$
for  $Q=4$ GeV/c at $p_{inc}=70$ GeV/c.
($\gamma $) in Fig.4: the calculated $A_N$ as
function of $x_F$ for inclusive lepton-pair production
using unpolarized proton-beam at 820 GeV/c and transversely polarized
nucleon and nuclear targets,
$p+p(\uparrow )\to l\bar l+X$,
$p+n(\uparrow )\to l\bar l+X$, and
$p+D(\uparrow )\to l\bar l+X$ for  $Q=4$ GeV/c.

Note that, while the qualitative results
(e.g. those shown in Table 1 and 2 and those for kaon-production listed above)
are obtained without any parameters, there is
one unknown parameter $C$ in all the qualitative results
(e.g. those shown in Figs. 2, 3 and 4).
This parameter $C$ characterizes the intensity of the surface
effect which ranges from 0 to 1.
We found $C=0.6$ by comparing the calculated curves with one
of the data points [4-7] of the reactions
$p(\uparrow )+p\to (\pi ^+,~\pi^0,~\pi^-)+X$ at
$p_{inc}= 200 $ GeV/c.

\section{Concluding remarks}

As far as singly polarized hadron-hadron
collisions are concerned, experimental
research is ahead of theoretical studies.
We theoriests need to work harder!
Our experimental colleagues can
help us not only by giving us more and better
data, but also by asking us more critical questions!

The characteristic features of the existing
single-spin asymmetry data show that they
have much in common with the typical properties
of soft hadronic processes observed in the
fragmentation regions of unpolarized hadron-hadron collisions.
These similarities strongly suggest that the
mechanism(s) responsible for such asymmetries are soft in nature.
Hence, it is not surprising to see that straight-forward
application of usual (leading-twist) perturbative QCD
leads to results in dramatic disagreement with the data.
The model (BRQM) proposed by the FU-Berlin group serves as
an example in which the relations between the
characteristic features of the existing data
and the non-pQCD aspects of such processes
are explicitly given.
In spite of its successful description of the
existing data and its prediction power,
BRQM is just a phenomenological model!
Whether --- if yes how --- this model can be embeded in the
framework of QCD --- the only candidate for hadronic interactions
--- is still an open question.
It is clear that
 the theorists have much home-work to do.
It is also clear that the
experimentalists in the Spin Physics Community have done a
magnificent job: Among other things,
they remind us theorists that it pays to be
open-minded and critical --- also towards our
favorite toy!


\newpage

\begin{table}[h]
\caption{Properties of $\pi^{\pm },\ \pi^0$ or $\eta $
in $p(\uparrow )+p\to \pi ^{\pm }({\rm or}\ \pi^0, \eta )+X$}
\centering
{\footnotesize
\begin{tabular}{l@{\extracolsep{0.6truecm}}
  c@{\extracolsep{0.6truecm}}c@{\extracolsep{0.6truecm}}
  c@{\extracolsep{0.6truecm}}c@{\extracolsep{1.0truecm}}
 l@{\extracolsep{0.6truecm}}
  c@{\extracolsep{0.6truecm}}c@{\extracolsep{0.6truecm}}
  c@{\extracolsep{0.6truecm}}
  c@{\extracolsep{0.6truecm}}c@{\extracolsep{0.6truecm}}
  c@{\extracolsep{0.6truecm}}}
\hline \\[-0.3truecm]
\multicolumn{5}{l} {$P$(sea)---$T$(val)} &
\multicolumn{7}{l} {$P$(val)---$T$(sea)} \\[-0.010truecm]
\hline \\[-0.3truecm]
$P$(sea)&$u$ &$\bar u$ &$d$ &$\bar d$ &
 $P$(val)&\multicolumn{2}{c}{$u$}      &  &\multicolumn{2}{c}{$d$}    &
 \\[-0.01truecm]
$p_y$   &0   &0        &0   &0        &
 $p_y$   &$\leftarrow$ &$\rightarrow $&  &$\leftarrow$&$\rightarrow$&
 \\[-0.01truecm]
Weight  &1   &1        &1   &1        &
 Weight  &5/3          &1/3           &  &1/3         &2/3          &

\\[-0.10truecm]
\hline\\[-0.3truecm]
$T$(val)& &$d$       &  &$u$      &
  $T$(sea)&\multicolumn{2}{c}{$\bar d$} &$d$&\multicolumn{2}{c}{$\bar u$}&$u$
  \\[-0.01truecm]
$p_y$   & &0         &  &0        &
  $p_y$   &\multicolumn{2}{c}{0}        &0  &\multicolumn{2}{c}{0}       &0
  \\[-0.01truecm]
Weight  & &1         &  &2        &
  Weight  &\multicolumn{2}{c}{1}        &1  &\multicolumn{2}{c}{1}       &1
  \\[-0.010truecm]
Product & &$d\bar u$ &  &$u\bar d$&
  Product &\multicolumn{2}{c}{$u\bar d$}&   &\multicolumn{2}{c}{$d\bar u$}&
  \\[-0.01truecm]
$p_y$   & &0         &  &0        &
  $p_y$   &$\leftarrow$ &$\rightarrow$ &   &$\leftarrow$ &$\rightarrow$ &
  \\[-0.01truecm]
Weight  & &1         &  &2        &
  Weight  &5/3          &1/3           &   &1/3          &2/3           &
  \\[-0.010truecm]
\hline\\[-0.3truecm]
$T$(val)& &$u$       &  &$d$      &
  $T$(sea)&\multicolumn{2}{c}{$\bar u$} &$u$&\multicolumn{2}{c}{$\bar d$}&$d$
  \\[-0.01truecm]
$p_y$   & &0         &  &0        &
  $p_y$   &\multicolumn{2}{c}{0}        &0  &\multicolumn{2}{c}{0}       &0
  \\[-0.01truecm]
Weight  & &2         &  &1        &
  Weight  &\multicolumn{2}{c}{1}        &1  &\multicolumn{2}{c}{1}       &1
  \\[-0.010truecm]
Product & &$u\bar u$ &  &$d\bar d$&
  Product &\multicolumn{2}{c}{$u\bar u$}&   &\multicolumn{2}{c}{$d\bar d$}&
  \\[-0.01truecm]
$p_y$   & &0         &  &0        &
  $p_y$   &$\leftarrow$ &$\rightarrow$ &   &$\leftarrow$ &$\rightarrow$ &
  \\[-0.01truecm]
Weight  & &2         &  &1        &
  Weight  &5/3          &1/3           &   &1/3          &2/3           &
  \\[-0.010truecm]\hline
\end{tabular}
   }
\label{tab:ANpp1}
\end{table}
\begin{table}[h]
\caption{Properties of $\pi^{\pm },\ \pi^0$ or $\eta $
in $\bar p(\uparrow )+p\to \pi ^{\pm }({\rm or}\ \pi^0, \eta )+X$}
\centering
{\footnotesize
\begin{tabular}{l@{\extracolsep{0.6truecm}}
  c@{\extracolsep{0.6truecm}}c@{\extracolsep{0.6truecm}}
  c@{\extracolsep{0.6truecm}}c@{\extracolsep{1.0truecm}}
 l@{\extracolsep{0.6truecm}}
  c@{\extracolsep{0.6truecm}}c@{\extracolsep{0.6truecm}}
  c@{\extracolsep{0.6truecm}}
  c@{\extracolsep{0.6truecm}}c@{\extracolsep{0.6truecm}}
  c@{\extracolsep{0.6truecm}}}
\hline \\[-0.3truecm]
\multicolumn{5}{l} {$P$(sea)---$T$(val)} &
\multicolumn{7}{l} {$P$(val)---$T$(sea)} \\[0.0truecm]
\hline \\[-0.36truecm]
$P$(sea)&$u$ &$\bar u$ &$d$ &$\bar d$ &
 $P$(val)&\multicolumn{2}{c}{$\bar u$}   &  &\multicolumn{2}{c}{$\bar d$}&
 \\[-0.05truecm]
$p_y$   &0   &0        &0   &0        &
 $p_y$   &$\leftarrow$ &$\rightarrow $&  &$\leftarrow$&$\rightarrow$     &
 \\[-0.01truecm]
Weight  &1   &1        &1   &1        &
 Weight  &5/3          &1/3           &  &1/3         &2/3          &
  \\[-0.10truecm]
  \hline\\[-0.35truecm]
$T$(val)& &$d$       &  &$u$      &
  $T$(sea)&\multicolumn{2}{c}{$d$}&$\bar d$&\multicolumn{2}{c}{$u$}&$\bar u$
  \\[-0.04truecm]
$p_y$   & &0         &  &0        &
  $p_y$   &\multicolumn{2}{c}{0}  &0       &\multicolumn{2}{c}{0}  &0
  \\[-0.01truecm]
Weight  & &1         &  &2        &
  Weight  &\multicolumn{2}{c}{1}  &1       &\multicolumn{2}{c}{1}  &1
  \\[-0.010truecm]
Product & &$d\bar u$ &  &$u\bar d$&
  Product &\multicolumn{2}{c}{$d\bar u$} & &\multicolumn{2}{c}{u$\bar d$}&
  \\[-0.01truecm]
$p_y$   & &0         &  &0        &
  $p_y$   &$\leftarrow$ &$\rightarrow$ &   &$\leftarrow$ &$\rightarrow$ &
  \\[-0.01truecm]
Weight  & &1         &  &2        &
  Weight  &5/3          &1/3           &   &1/3          &2/3           &
  \\[-0.010truecm]
\hline\\[-0.30truecm]
$T$(val)& &$u$       &  &$d$      &
  $T$(sea)&\multicolumn{2}{c}{$u$}&$\bar u$&\multicolumn{2}{c}{$d$}&$\bar d$
  \\[-0.01truecm]
$p_y$   & &0         &  &0        &
  $p_y$   &\multicolumn{2}{c}{0}        &0  &\multicolumn{2}{c}{0}       &0
  \\[-0.01truecm]
Weight  & &2         &  &1        &
  Weight  &\multicolumn{2}{c}{1}        &1  &\multicolumn{2}{c}{1}       &1
  \\[-0.010truecm]
Product & &$u\bar u$ &  &$d\bar d$&
  Product &\multicolumn{2}{c}{$u\bar u$}&   &\multicolumn{2}{c}{$d\bar d$}&
  \\[-0.01truecm]
$p_y$   & &0         &  &0        &
  $p_y$   &$\leftarrow$ &$\rightarrow$ &   &$\leftarrow$ &$\rightarrow$ &
  \\[-0.01truecm]
Weight  & &2         &  &1        &
  Weight  &5/3          &1/3           &   &1/3          &2/3           &
  \\[-0.010truecm]\hline
\end{tabular}
   }
\label{tab:ANpbarp}
\end{table}

\newpage

\begin{center}
\mbox{\epsfig{file=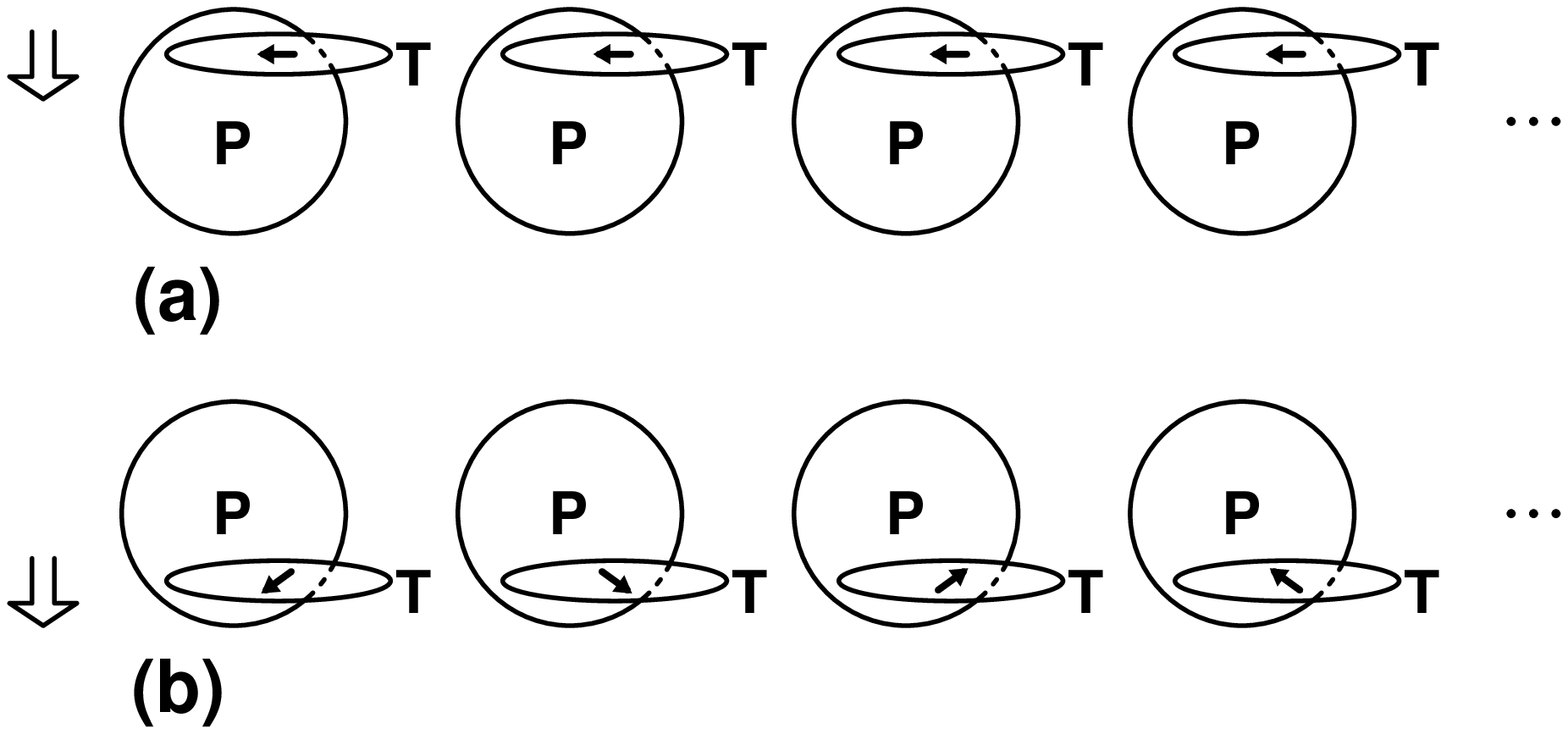,height=12cm,angle=270}}
\end{center}
\vskip -1.6truecm
\noindent
{\large\sf Figure~1:}~

\vskip 0.5truecm

\begin{center}
\mbox{\epsfig{file=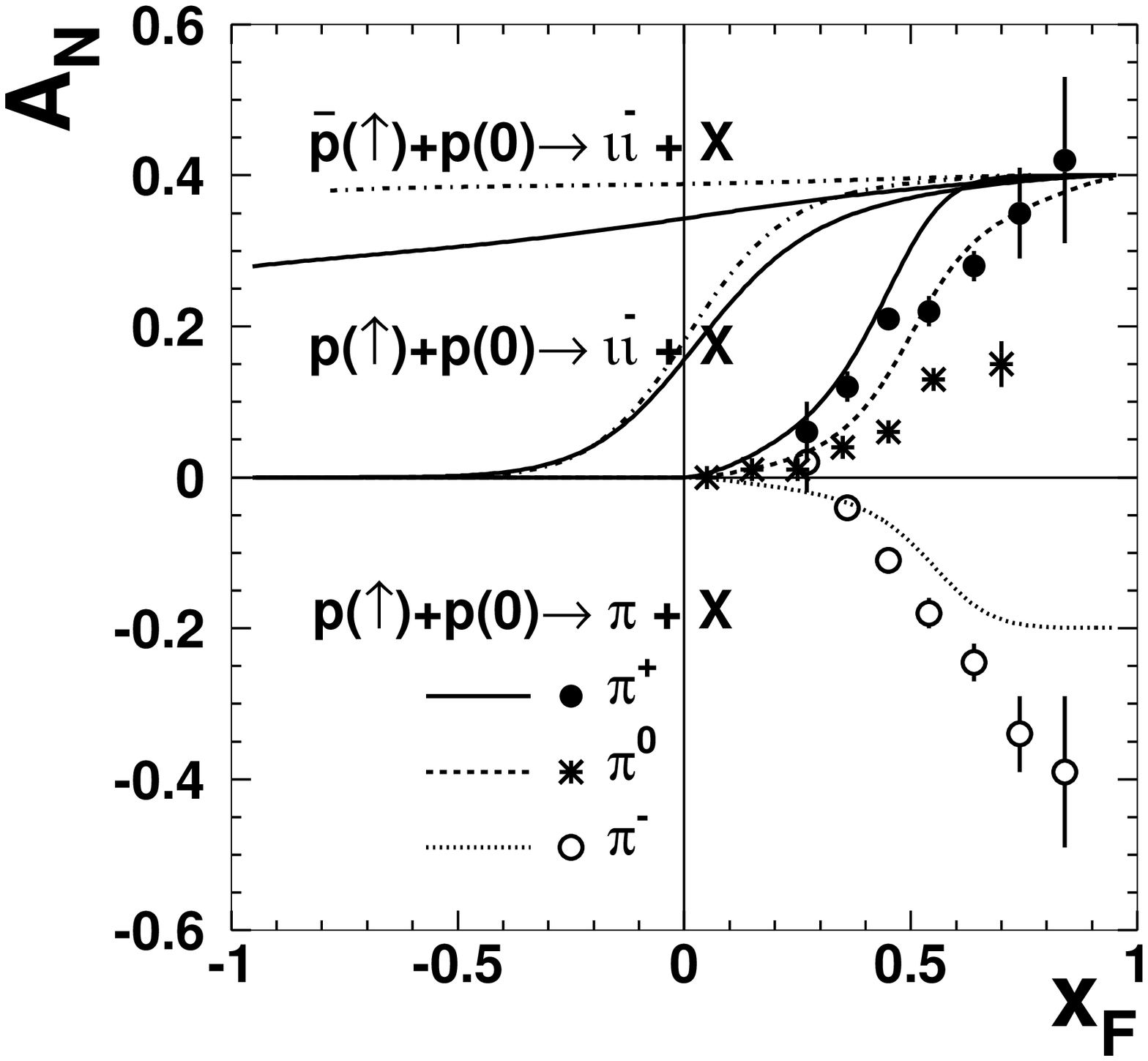,height=10cm}}
\end{center}
\vskip -0.9truecm
\noindent
{\large\sf Figure~2:}~

\vskip 0.5truecm

\begin{center}
\mbox{\epsfig{file=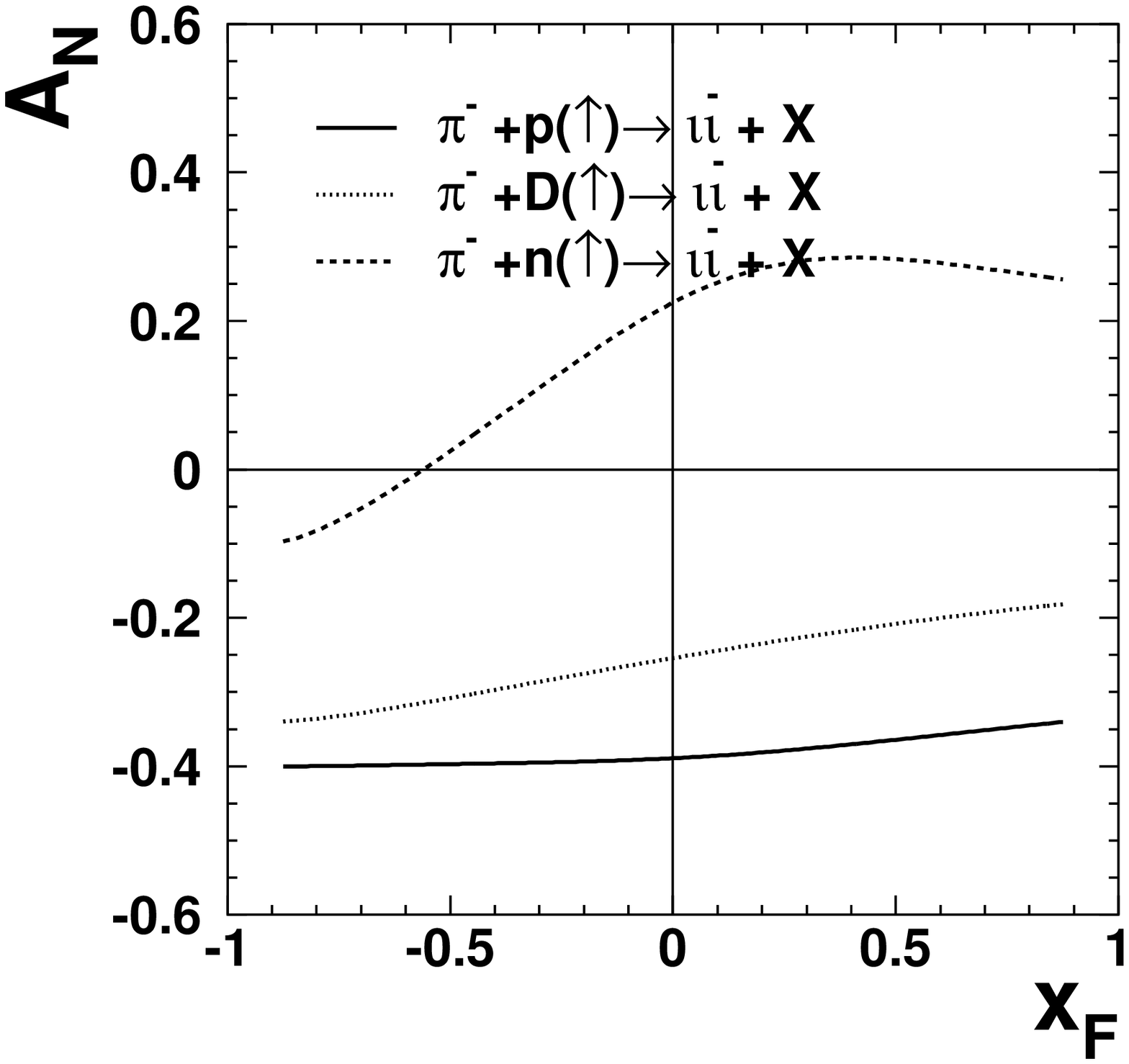,height=10cm}}
\end{center}
\vskip -0.9truecm
\noindent
{\large\sf Figure~3:}~

\vskip 0.5truecm

\begin{center}
\mbox{\epsfig{file=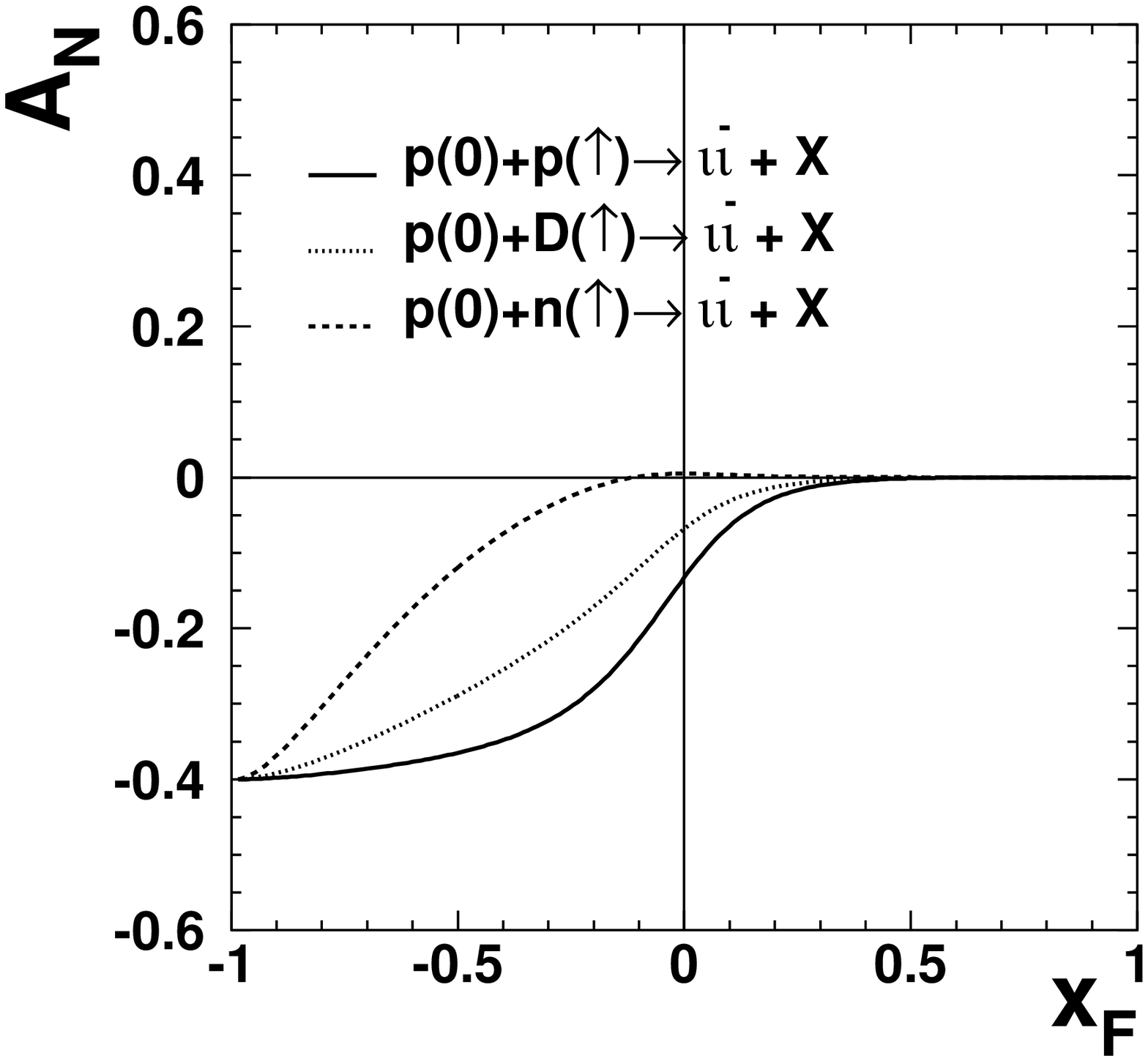,height=10cm}}
\end{center}
\vskip -0.9truecm
\noindent
{\large\sf Figure~4:}~


\begin{thebibliography}{999}
%
%
\bibitem{ref1} See, for example, \\
  P.R. Cameron et al., Phys. Rev. {\bf D32},3070 (1985);\\
  A.D. Krisch, in
  Proceedings of the 9th International Symposium
  on High-Energy Spin Phys., 1990,
  Bonn, ed. K.H. Althoff and W. Meyer, Springer Verlag (1991), p.20;
  and the references cited therein.
\bibitem{ref2} S.~Saroff et al., Phys. Rev. Lett. {\bf 64}, 995 (1990);
  and the references cited therein.
\bibitem{ref3} V.D. Apokin et al., Phys. Lett. {\bf B243}, 461 (1990);
  and "$x_F$-dependence of the asymmetry in inclusive $\pi ^0$ and
  $\eta ^0$ production in beam fragmentation region",
  Serpuhkov-Preprint (1991).
\bibitem{e704a} FNAL E581/704 Collaboration,
   D.L. Adams et al., Phys. Lett. {\bf B261}, 201 (1991).
\bibitem{e704b} FNAL E704 Collaboration,
  D.L.~Adams et al.,
  Phys. Lett. {\bf B264}, 462 (1991).
\bibitem{e704c} FNAL E704 Collaboration,
  D.L.~Adams et al.,
  Phys. Lett.  {\bf B276}, 531 (1992).
\bibitem{e704d} FNAL E704 Collaboration,
  D.L.~Adams et al.,
  Z. Phys. {\bf C56},181 (1992);
\bibitem{e704e}
 A. Yokosawa, in
 {\it Frontiers of High Energy Spin Physics},
 Proceedings of the 10th
 International Symposium, Nagoya, Japan 1992,
 edited by T. Hasegawa {\it et al.}
 (Universal Academy, Tokyo, 1993),
 and the references given therein.
\bibitem{9}
 S.~Nurushev, in
 Proceedings of the 11th
 International Symposium on High Energy Spin Physics,
 Bloomington, Indiana, 1994,
 and the references given therein.
\bibitem{10} FNAL E704 Collaboration, D.L.~Adams et al., Phys. Lett.
 {\bf B 345}, 569 (1995).
\bibitem{11} G.~Kane, J.~Pumplin and W.~Repko,
   Phys. Rev. Lett. {\bf 41}, 1689 (1978).
\bibitem{12} D. Sivers, Phys. Rev. {\bf D41},83 (1990).
\bibitem{13} D. Sivers, Phys. Rev.  {\bf D41}, 261 (1991).
\bibitem{14} J.~Qiu and G.~Sterman, Phys. Rev. Lett. {\bf 67}, 2264 (1991).
\bibitem{15} A.~Sch\"afer, L.~Mankiewicz, P.~Gornicki and
   S.~G\"ullenstern, Phys. Rev. {\bf D47}, 1 (1993).
\bibitem{16} J.~Collins, Nucl. Phys. {\bf 396}, 161 (1993).
\bibitem{17} A.V.~Efremov, V.M.~Korotkiyan, and O.V.~Teryaev,
 Phys. Lett. {\bf B 348}, 577 (1995).
\bibitem{18} M.~Anselmino, M.~Boglione, and F.~Murgia,
  Preprint INFNCA-TH-94-27 (revised version) (1995).
\bibitem{19} G.~Belletini et al., Phys. Lett. {\bf 45B}, 69 (1973).
\bibitem{20} W.~Ochs, Nucl. Phys. {\bf B118}, 397 (1977).
\bibitem{21} K.P. Das and R.C. Hwa, Phys. Lett. {\bf 68B}, 459 (1977).
\bibitem{22} R.C. Hwa, Z. Phys. {\bf C20}, 27 (1983).
\bibitem{23} M.G. Albrow et al., Nucl. Phys. {\bf B51}, 388 (1973).
\bibitem{24} G. Giacomelli and M. Jacob, Phys. Rep. {\bf 55}, 38 (1979).
\bibitem{25} See, for example, J. Franklin, in
 {\it Proceedings of the 8th International Symposium on
 High Energy Spin Physics}, Minneapolis, Minnesota, 1988,
 edited by K.~Heller, AIP Conf. Proc. No. 187 (AIP, New York, 1989),
 pp.298 and 384.
\bibitem{26} T. Meng, in Proc. of the 4th Workshop on
  High Energy Spin Physics,
  Protvino, Russia, Sept. 1991, ed. S.B. Nurushev, p. 121 (1991).
\bibitem{re27} Z. Liang and T. Meng, Z. Phys. {\bf A344}, 171 (1992).
\bibitem{re28} C. Boros, Z. Liang and T. Meng, Phys. Rev. Lett. {\bf 70}, 1751
(1993).
\bibitem{re29} Z. Liang and T. Meng, Phys. Rev. {\bf D49}, 3759 (1994).
\bibitem{re30} C. Boros, Z. Liang and T. Meng, Phys. Rev.
 {\bf D51}, 4698 (1995).
\bibitem{re31} C. Boros and T. Meng, Phys. Rev. {\bf D52}, 529 (1995).
\bibitem{re32} Z. Liang, T. Meng and R. Rittel, Phys. Rev. {\bf D} (submitted).
\end{thebibliography}
\end{document}